\documentclass[aps,prb,twocolumn,superscriptaddress,showpacs,preprintnumbers,amsmath,amssymb,floatfix]{revtex4}
\usepackage{graphicx}
\usepackage{dcolumn}
\usepackage{bm}
\usepackage{color}
\usepackage[latin1]{inputenc}

\begin{document}

\title{Wave-packet scattering at a normal-superconductor interface in two-dimensional materials: a generalized theoretical approach}

\author{F. J. A. Linard}
\affiliation{Departamento de F\'{\i}sica, Universidade Federal do Cear\'a, Caixa Postal 6030, 60455-760 Fortaleza, Cear\'a, Brazil}
\author{V. N. Moura}
\affiliation{Departamento de F\'{\i}sica, Universidade Federal do Cear\'a, Caixa Postal 6030, 60455-760 Fortaleza, Cear\'a, Brazil}
\author{L. Covaci}
\affiliation{Department of Physics, University of Antwerp, Groenenborgerlaan 171, B-2020, Antwerpen, Belgium}
\author{M. V. Milo\v{s}evi\'c}
\affiliation{Department of Physics, University of Antwerp, Groenenborgerlaan 171, B-2020, Antwerpen, Belgium}
\author{A. Chaves} \email{andrey@fisica.ufc.br}
\affiliation{Departamento de F\'{\i}sica, Universidade Federal do Cear\'a, Caixa Postal 6030, 60455-760 Fortaleza, Cear\'a, Brazil}
\affiliation{Department of Physics, University of Antwerp, Groenenborgerlaan 171, B-2020, Antwerpen, Belgium}

\begin{abstract}
A wave-packet time evolution method, based on the split-operator technique, is developed to investigate the scattering of quasi-particles at a normal-superconductor interface of arbitrary profile and shape. As a practical application, we consider a system where low energy electrons can be described as Dirac particles, which is the case for most two-dimensional materials, such as graphene and transition metal dichalcogenides. However the method is easily adapted for other cases such as electrons in few layer black phosphorus, or any Schr\"odinger quasi-particles within the effective mass approximation in semiconductors. We employ the method to revisit Andreev reflection in graphene, where specular and retro reflection cases are observed for electrons scattered by a step-like superconducting region. The effect of opening a zero-gap channel across the superconducting region on the electron and hole scattering is also addressed, as an example of the versatility of the technique proposed here.
\end{abstract}

\maketitle

\section{Introduction}

It is widely known that electron states convert to holes after being reflected by a normal (N)/superconductor (SC) interface. \cite{Andreev} This effect, also known as Andreev reflection, exhibits peculiarities: if the incidence to the NS interface is normal, the electron is fully converted into a hole, whereas for oblique incidence, part of the wave function is reflected back to the normal region as an electron state. In a system consisting of a semiconductor material, with a considerable energy gap separating conduction and valence bands, the momentum of the hole, along with its energy dispersion, guarantees that the hole component of the wave function travels back in a trajectory that is parallel to that of the incident electron, which is then coined the term retro-reflection. However, it has been demonstrated that in monolayer graphene, where low energy electrons behave as massless Dirac fermions in a gapless band structure, \cite{GrapheneReview} the energy dispersion is such that, for low Fermi levels, the hole component of the wave function travels back in the normal region in a trajectory that is parallel to that of the \textit{reflected} electron, thus undergoing a \textit{specular} Andreev reflection. This effect has been predicted by Beenakker in 2006 \cite{Beenakker} in a model for monolayer graphene, which was further extended to bilayer graphene, \cite{Ludwig, Takane} and experimentally observed only very recently \cite{Efetov1, Efetov2, Takane, Soori}

Further suggestions have been made for experimental observation of Andreev scattering using N/SC interfaces based on different materials, such as transition metal dichalcogenides \cite{Bai2020, Yang2016} and their heterostructures, \cite{Habe2019} as well as on borophene. \cite{Zhou2020} A NS interface in monolayer black phosphorus \cite{BlackPhosphorus} has also been recently theoretically proposed as a venue for the observation of Andreev reflection. \cite{Linder} Since this is a $\approx$ 2 eV gap semiconductor, \cite{CastellanosGomez} only retro reflection is expected to occur, but many two-dimensional (2D) materials with zero gap exist \cite{Das2015, PolaritonsReview, AndreyReview} and may be suitable for the observation of specular Andreev reflection too. However, developing a different theory for each Hamiltonian describing each of the several classes of materials in the 2D materials family seems like an insurmountable challenge. Moreover, most of the techniques proposed in the literature for the study of Andreev scattering resort to plane waves-based methods which, although providing analytical solutions to the scattering problem, are harder to be adapted to physical situations involving arbitrary potentials and N/SC interface profiles, as well as in the presence of applied fields. This motivates us to develop a method that is easily adapted for any configuration of the potential and N/SC interface profiles, as well as for any form of the Hamiltonian describing the materials involved.             

In this paper, we develop a numerical technique to investigate wave-packet dynamics at an N/SC interface, based on an extension of the so called split-operator method, \cite{AndreySplitOperator, DeganiReview} that accounts for the Bogoliubov-de Gennes Hamiltonian describing a superconductor. The method allows for the investigation of wave-packet scattering at the interface and the interplay between electron and hole states, allowing one to assume an arbitrary form for the interface and potential profiles and to conveniently change the system Hamiltonian for that of any 2D material. We apply the method to calculate transmission probabilities in a system consisting of a normal wave-guide defined by adjacent superconducting regions. Our results illustrate how the channel width and length can be used to tune the electron and hole components of the wave-packet that leaves the channel region. 

\section{wave-packet propagation method}

Consider a basis $(u_A ~ u_B ~ v_A ~ v_B)^T$, where $u_i$ and $v_i$ ($i = A, B$) represent the $i$-th component of the 2-component spinor describing electrons and holes, respectively. The Bogoliubov-de Gennes (BdG) Hamiltonian \cite{Bogoliubov} describing the NS interface is given by
\begin{equation}
H_{BdG} = \left( \begin{array}{cc}
H-E_F + U(\vec r) & \Delta(\vec r)\\
\Delta^{*}(\vec{r}) & -[H-E_F +U(\vec r)]
\end{array} \right),
\end{equation}
where $H$ is a 2$\times$2 matrix Hamiltonian for charged particles in the material in its normal phase, $\Delta(\vec r) =\Delta_0(\vec{r})e^{i\phi}$ is a space-dependent superconducting gap, which is assumed to be non-zero only at the superconducting region, $U(\vec{r})$ is an external potential, and $E_F$ is the Fermi level. Notice that each $U$, $\Delta$, and $E_F$ must be multiplied by a 2$\times$2 identity matrix $\mathcal{I}$ (omitted here for the sake of convenience), so that $H_{BdG}$ is a 4$\times$4 matrix.

The time evolution of an arbitrary initial wave-packet
\begin{equation}\label{eq.wavepacket}
\Psi(\vec r,t = 0) = \left(\begin{array}{c}
u_A\\
u_B\\
v_A\\
v_B
\end{array}\right)\times \psi(\vec r, 0),
\end{equation}
is calculated as
\begin{equation}\label{eq.timeevol0}
|\Psi(\vec{r},t+\Delta t)\rangle = e^{-i\frac{H_{BdG}}{\hbar}\Delta t} | \Psi(\vec{r},t) \rangle.
\end{equation}
The Hamiltonian $H_{BdG}$ is conveniently split into parts that depend exclusively on real or reciprocal space coordinates
\begin{eqnarray}
H_{BdG} = (H-E_F)\otimes \sigma_z + U(\vec r)\otimes\sigma_z \nonumber\\
+\Delta_0(\vec{r})(\cos\phi \mathcal{I}\otimes \sigma_x + \sin\phi \mathcal{I} \otimes \sigma_y),
\end{eqnarray}
where the first term retains only the terms that depend on reciprocal-space coordinates $\vec{k}$ and $\vec{\sigma}$ is the vector of Pauli matrices. The Suzuki-Trotter expansion\cite{Suzuki} of the exponential in the time evolution operator in Eq. (\ref{eq.timeevol0}) yields 
\begin{equation}\label{eq.timevol}
e^{-i\frac{H_{BdG}}{\hbar}\Delta t} = e^{-i \vec{W}_r \otimes \vec{\sigma}}e^{-i\vec{W}_k \otimes \vec{\sigma} }e^{-i\vec{W}_r \otimes \vec{\sigma}} 
+ O(\Delta t^3)
\end{equation}
where $\vec{W_r} = \left(\Delta_0\cos\phi,\Delta_0\sin\phi, U\right)\Delta t/2\hbar$, $\vec{W_k}$ = $(0,0,H-E_F)\frac{\Delta t}{\hbar}$, and the $O(\Delta t^3)$ error comes from the non-commutativity between $\vec{W}_r \cdot \vec{\sigma}$ and $\vec{W}_k \cdot \vec{\sigma}$ operators. Since\cite{AndreySplitOperator}
\begin{equation}\label{eq.splitoperatorspin}
e^{-i \vec{S} \cdot \vec{\sigma}} = \left( \begin{array}{cc}
\cos(S) - i \sin(S)\frac{S_z}{S} & - i \sin(S)\frac{S_x - i S_y}{S}\\
- i \sin(S)\frac{S_x + i S_y}{S} & \cos(S) + i \sin(S)\frac{S_z}{S}
\end{array}\right),
\end{equation}
each of the exponentials in Eq. (\ref{eq.timevol}) are expanded in an exact way. This approach will be demonstrated to be very convenient in the context of 2D materials, since low energy electrons in these systems are often described by 2$\times$2 Hamiltonians that can be re-written into the form $H = \vec{h} \cdot \vec{\sigma}$, provided one considers a proper $\vec{h}$.\cite{Bieniek2018, Cunha2020}

Using Eq. (\ref{eq.splitoperatorspin}) to expand the exponentials in Eq. (\ref{eq.timevol}), one obtains the final form of the time evolution operator as a series of multiplications between 4 $\times$ 4 matrices given by
\begin{eqnarray}\label{eq.Splitr}
e^{-i \vec{W_r} \otimes \vec{\sigma}} = \mathcal{M}_r = \left( \begin{array}{cccc}
A_{-} & 0 & B_{-} & 0 \\
0 & A_{-} & 0 & B_{-}\\
B_{+} & 0 & A_{+} & 0\\
0 & B_{+} & 0 & A_{+}
\end{array}\right)
\end{eqnarray}
where $A_{\pm} = \cos\left ( \frac{\Delta t}{2 \hbar} \sqrt{\Delta_0^{2}+U^2} \right ) \pm i \sin\left ( \frac{\Delta t}{2 \hbar} \sqrt{\Delta_0^{2}+U^2} \right )\frac{U}{\sqrt{\Delta_0^{2}+U^2}}$ and $B_{\pm} = -i \sin\left ( \frac{\Delta t}{2 \hbar} \sqrt{\Delta_0^{2}+U^2} \right )\frac{\Delta_0 e^{\pm i\phi}}{\sqrt{\Delta_0^{2}+U^2}}$; and
\begin{eqnarray}
\label{eq.Splitk} e^{-i \vec{W_k} \otimes \vec{\sigma}} = \mathcal{M}_k = \left( \begin{array}{cccc}
C_{-} & D_{-} & 0 & 0\\
D_{+} & C_{+} & 0 & 0\\
0 & 0 & C_{-}^{'} & D_{-}^{'}\\
0 & 0 & D_{+}^{'} & C_{+}^{'}
\end{array}\right),
\end{eqnarray}
where $C_{\pm} = \left ( \cos(\omega) \pm i \sin(\omega)\frac{\omega_z}{\omega} \right )e^{iE_F\frac{\Delta t}{\hbar}}$, $D_{\pm} = -i\sin(\omega)\frac{\omega_x \pm i \omega_y}{\omega}e^{iE_F\frac{\Delta t}{\hbar}}$, $C_{\pm}^{'} = \left ( \cos(\omega') \pm i\sin(\omega')\frac{\omega'_z}{\omega'} \right )e^{-iE_F\frac{\Delta t}{\hbar}}$, $D_{\pm}^{'} = -i\sin(\omega')\frac{\omega'_x \pm i \omega'_y}{\omega'}e^{-iE_F\frac{\Delta t}{\hbar}}$, $\vec{\omega} = \left(h_x,h_y,h_z\right)\frac{\Delta t}{\hbar}$ and $\vec{\omega'} = \left(-h_x,-h_y,-h_z\right)\frac{\Delta t}{\hbar}$. 

\begin{figure}[!t]
\centerline{\includegraphics[width = \linewidth]{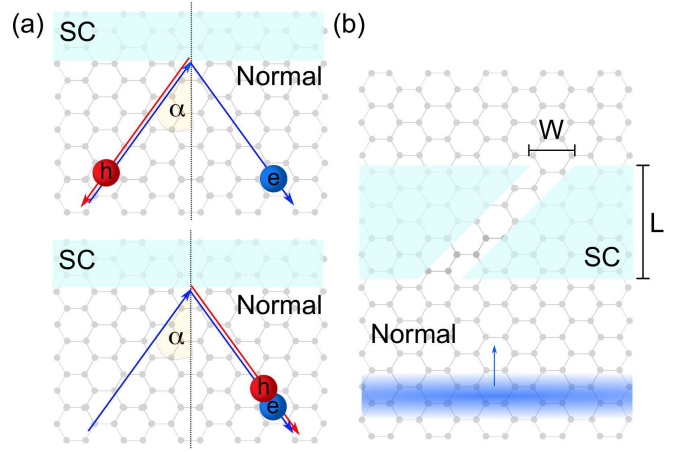}}
\caption{(Color online) Sketch of the two graphene-based systems considered here: (a) a single interface between normal and superconducting (SC) regions, and (b) a tilted (by 45$^o$) channel of length $L$ and width $W$ across the SC region. In the former, Andreev retro-(top) and specular (bottom) reflections will be investigated by calculating the trajectories of electron (e, blue) and holes (h, red), assuming an incidence angle $\alpha$ and describing the quasi-particles as circular gaussian wave-packets. As for the latter, we will investigate transmission/reflection probabilities for an incoming electron described by a gaussian wave front (blue gradient). } \label{fig:Sketch}
\end{figure}

Thus, a wave-packet at an instant $t$, $|\Psi(\vec{r},t)\rangle$, is propagated to $|\Psi(\vec{r},t+\Delta t)\rangle$ as
\begin{equation}\label{eq.timeevolfinal}
|\Psi(\vec{r},t+\Delta t)\rangle = \mathcal{M}_r \mathcal{M}_k \mathcal{M}_r  | \Psi(\vec{r},t) \rangle,
\end{equation}
which is performed in three steps: (i) multiplying $|\Psi(\vec{r},t)\rangle$ by $\mathcal{M}_r$, (ii) taking the Fourier transform of the resulting spinor and multiplying it by $\mathcal{M}_k$ in reciprocal space, and then (iii) taking the resulting spinor back to real space, by performing an inverse Fourier transform on it, and multiplying it by $\mathcal{M}_r$ again. The process is repeated until the propagation is performed for a given time interval. Notice that, since the matrix expansion in Eq. (\ref{eq.splitoperatorspin}) is exact, the only error involved in this procedure is the $O(\Delta t^3)$ error resulting from the Suzuki-Trotter expansion in Eq. (\ref{eq.timevol}). As we consider a small time step $\Delta t = 0.1$ fs, this term can be neglected from now on. 

Electron and hole probability densities are calculated from the propagated electron-hole pseudo-spinor
\begin{equation}
\Psi(\vec r,t) = \left(\begin{array}{c}
\psi_{uA}(\vec r, t)\\
\psi_{uB}(\vec r, t)\\
\psi_{vA}(\vec r, t)\\
\psi_{vB}(\vec r, t)
\end{array}\right)
\end{equation}
as 
\begin{eqnarray}
P_e (t) = \int_{r_1}^{r_2} [|\psi_{uA}(\vec r, t)|^2 + |\psi_{uB}(\vec r, t)|^2]d \vec r \label{eq.Pe}\\
P_h (t) = \int_{r_1}^{r_2} [|\psi_{vA}(\vec r, t)|^2 + |\psi_{vB}(\vec r, t)|^2]d \vec r , \label{eq.Ph}
\end{eqnarray}
where the interval $[r_1,r_2]$ limits the region of interest in space. Reflection (transmission) probabilities are obtained as the converged values of Eqs. (\ref{eq.Pe}) and (\ref{eq.Ph}), integrated only within the space before (after) the SC region, as $t \rightarrow \infty$.

\section{Results}

\subsection{Uniform normal-SC interface in Dirac-Weyl materials: revisiting Andreev reflection in graphene}

Let us first revisit the problem of Andreev reflection in graphene. Figure 
\ref{fig:Sketch}(a) shows a sketch of the proposed situation, where an electron in normal graphene propagates towards the superconducting region (shaded) through a trajectory that makes an angle $\alpha$ with the direction normal to the interface. 

For the envelope function multiplying the pseudo-spin in Eq. (\ref{eq.wavepacket}), we assume a gaussian wave-packet
\begin{equation}\label{eq.packet}
\psi(\vec r, 0) = \frac{1}{d\sqrt{2\pi}} \exp\left[{-\frac{(x-x_0)^2+(y-y_0)^2}{2d^2}+i\vec{k}_0\cdot\vec{r}}\right]
\end{equation}
describing a propagating low energy electron in graphene. The band structure of Dirac-Weyl materials (e.g. graphene) around K and K' points of the first Brillouin zone can be approximated by linear functions that follow from diagonalization of the effective Hamiltonian
\begin{equation}\label{eq.HamPlusMinus}
H_{\pm} = \hbar v_F (\pm k_x \sigma_x + k_y \sigma_y),
\end{equation}
where $v_F$ is the Fermi velocity and $\pm$ refers to K(+) and K'(-) cones, so that low energy electrons in this material behave as massless Dirac fermions. These cones are related by time-reversal symmetry, therefore, here we will consider only the case of electrons around K, whereas the behavior of electrons at K' are predicted from our results just by applying straighforward transformations due to the sign change in Eq. (\ref{eq.HamPlusMinus}). This Hamiltonian enters Eq. (\ref{eq.Splitk}) through the $\vec{\omega} = \vec{h}\Delta t/\hbar$ and $\vec{\omega'} = -\vec{h}\Delta t/\hbar$ terms, in this case, constructed by re-writting $H_{\pm} = (\pm h_x, h_y, 0)\cdot \vec{\sigma}$ with $\vec{h} = \hbar v_F\vec{k}$. The calculation is easily adapted e.g. for bilayer and trilayer graphene (in the ABC stacking order), using the 2$\times$2 approximation for the Hamiltonian proposed in Ref. [\onlinecite{McCann}], where one just needs to re-define $\vec{h} = \frac{\hbar^2 v_F^2}{\gamma}(k_x^2+k_y^2,\pm 2k_xk_y,0)$ and $\vec{h} = \frac{\hbar^3 v_F^3}{\gamma^2}(k_x^3-k_y^2k_x,3k_x^2k_y-k_y^3,0)$ for bilayer and trilayer cases,\cite{Lavor2020} respectively, with $\gamma$ as the inter-layer hopping parameter.

The external potential is taken as $U(\vec{r})\equiv 0$ and the superconducting gap $\Delta(\vec{r})$ is assumed to be a step functions that is zero for $y \leq 0$ and $\Delta_0$ otherwise. We also assume a zero superconducting phase $\phi = 0$. From now onwards, we write energies in units of the SC gap $\Delta_0$ and spatial coordinates in units of $r_0 = \hbar v_F / \Delta_0$. The wave-packet energy is fixed as $\epsilon = 0.7 \Delta_0$, which is used as input for Eq. (\ref{eq.packet}) through the modulus of the wave vector $k_0 = (\epsilon + E_F)/\Delta_0 r_0$. The wave-packet width is fixed as $d = 6.67 r_0$, which represents a $\Delta E \approx 0.15 \Delta_0$ width in energy space.  

\begin{figure}[!t]
\centerline{\includegraphics[width = \linewidth]{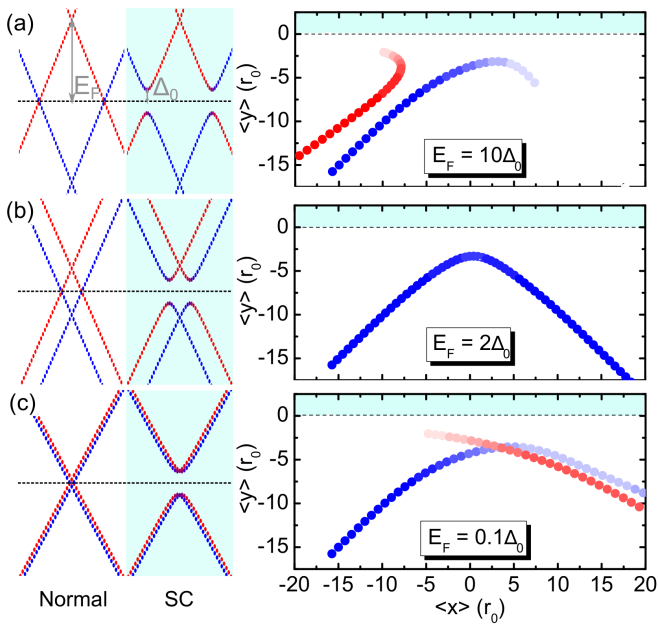}}
\caption{(Color online) Sketch of the band structures in the normal and SC regions (left), and wave-packet trajectories (right), considering an incidence angle $\alpha = 45^o$ and three values of Fermi energy: (a) $E_F = 10 \Delta_0$, (b) $E_F = 2 \Delta_0$, and (a) $E_F = 0.1 \Delta_0$. SC regions are highlighted as shaded areas in the figures. Color map in right panels is such that {\color{red}{red}} ({\color{blue}{blue}}) represents hole (electron) wave-packets, whereas darker colors represent higher probability densities.} \label{fig:ExW}
\end{figure}

Figure \ref{fig:ExW} sketches the band diagrams in the normal (white) and superconducting (shaded) regions, for different values of Fermi level. When the Fermi level is much larger than the superconducting gap, $E_F\gg \Delta_0$, electrons with energy $\epsilon < \Delta_0$ inciding in the superconducting region are reflected partially as holes. If the incidence is normal, the electron-hole conversion occurs with unit probability. On the other hand, if the trajectory of the incident electron makes a non-zero angle $\alpha$ with the vertical axis (see trajectories in Fig. \ref{fig:ExW} for $\alpha = 45^o$), a normal (electron) reflection is also expected. Moreover, the reflected hole is expected to propagate along the same trajectory as the incident electron, but with opposite propagation direction, which is known as Andreev retro-reflection. This is verified in the trajectory of electrons (blue symbols) and holes (red symbols) in Fig. \ref{fig:ExW}(a), where darker (brighter) colors represent higher (lower) probability density. A small Goos-H\"anchen shift is also observed between electron and hole trajectories. \cite{GoosHanchen} The picture is however different if $E_F$ is in the same order of magnitude as $\Delta_0$, as in Fig. \ref{fig:ExW}(b), where the almost no electron-hole conversion is observed. Furthermore, if $E_F\ll \Delta_0$, the converted hole wave function propagates in the same direction as the reflected electron, as one verifies in \ref{fig:ExW}(c), which is known as Andreev specular reflection.

\begin{figure}[!t]
\centerline{\includegraphics[width = \linewidth]{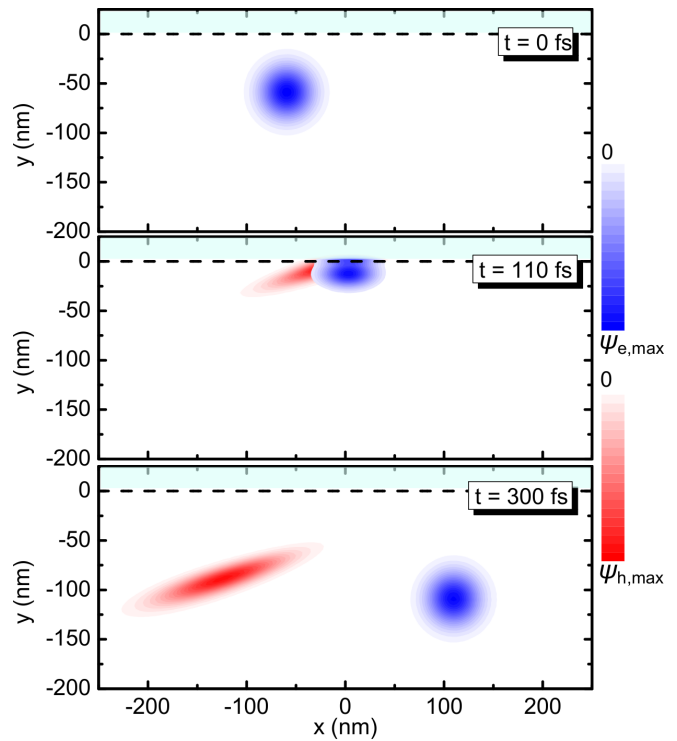}}
\caption{(Color online) Snapshots of the wave-packet projections over the electron (blue) and hole (red) states, for a wave-packet in graphene being reflected by a superconduction region (shaded area) in $y > 0$, assuming $E_F = 10 \Delta_0$. The snapshots are taken at three different instants in time, namely, $t = 0, 110$ fs and 300 fs.} \label{fig:SnapShots}
\end{figure}

\begin{figure}[!t]
\centerline{\includegraphics[width = \linewidth]{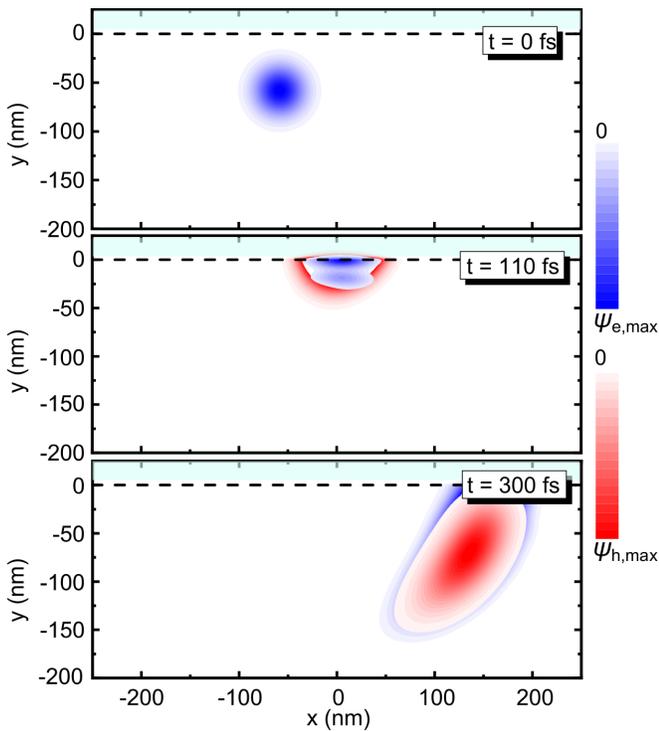}}
\caption{(Color online) The same as \ref{fig:SnapShots}, but for $E_F = 0.1 \Delta_0$.} \label{fig:SnapShots01}
\end{figure}

Figures \ref{fig:SnapShots} and \ref{fig:SnapShots01} illustrate Andreev retro- and specular reflections, respectivelly, by showing snapshots of the electron (blue) and hole (red) probability density distributions at three different instants in time. In the former (latter) the scattered electron and hole wave-packets clearly propagate towards opposite (the same) directions.

\begin{figure}[!t]
\centerline{\includegraphics[width = \linewidth]{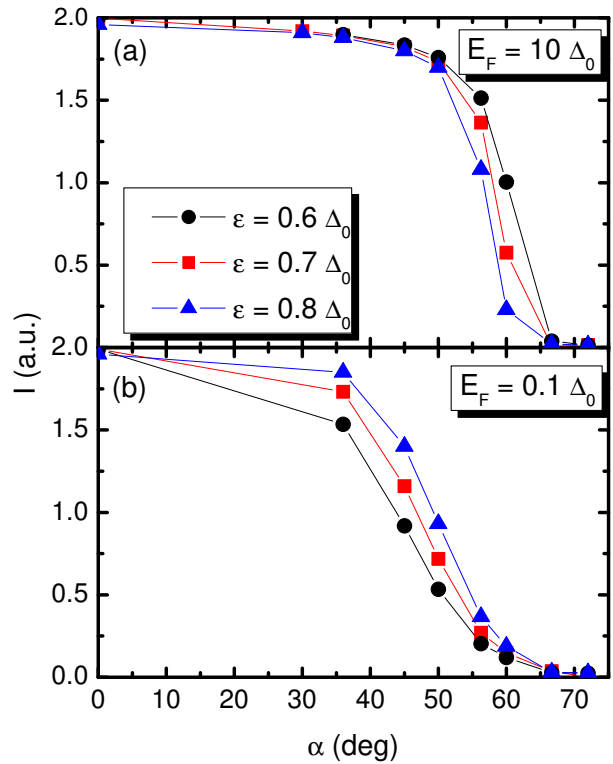}}
\caption{(Color online) Integration kernel in the BTK model of conductivity as a function of the wave-packet incidence angle, assuming wave-packets with different energies $\varepsilon$ and Fermi levels (a) $E_F = 10 \Delta_0$ and (b) $E_F = 0.1 \Delta_0$.} \label{fig:IKernel}
\end{figure}

Within the Blonder-Tinkham-Klapwijk (BTK) model, conductivity is proportional to $\int_0^{\infty}[1-r(\varepsilon,\alpha)+r_A(\varepsilon,\alpha)]\cos \alpha d\alpha$, where $r$ and $r_A$ represent probabilities of observing a reflected electron and hole, respectively, after scattering of the incident electron by the SC interface. In graphene normal-SC interface, it is known that in the case of retro-(specular)reflection, i.e. for $E_F > \Delta_0$($E_F < \Delta_0$), increasing the voltage $V$ leads to an increase (decrease) in the conductivity.\cite{Beenakker} It is not in the scopus of this paper to calculate the exact value of the conductivity. Nevertheless, one can use the method proposed here to verify this result. The integration kernel $I(\varepsilon,\alpha) = [1-r(\varepsilon,\alpha)+r_A(\varepsilon,\alpha)]$ in the BTK expression is plotted as a function of the incidence angle in Fig. \ref{fig:IKernel}, assuming two values of Fermi level. Increasing the energy of the incident wave-packet, which plays the role of the voltage $V$ in BTK model, leads to $I\times \alpha$ curves with consistently larger area when $E_F > \Delta_0$, as in Fig. \ref{fig:IKernel}(a). Consequently, the integral of $I$ with respect to the angle $\alpha$ increases with $\varepsilon$, thus suggesting a conductivity that increases with $V$. The opposite is true for $E_F < \Delta_0$, as in Fig. \ref{fig:IKernel}(b), where increasing the wave-packet energy rather decreases the area of the $I\times \alpha$ and, consequently, the conductivity.           

\subsection{Zero-gap channel in the superconducting region}

We now investigate the propagation of a wave front across a channel open in the SC region, as illustrated in Fig. \ref{fig:Sketch}(b). The channel is tilted by 45$^o$ from the vertical axis, so that the first reflection by the normal-SC interface makes the electron propagate horizontally.

\begin{figure}[!t]
\centerline{\includegraphics[width = \linewidth]{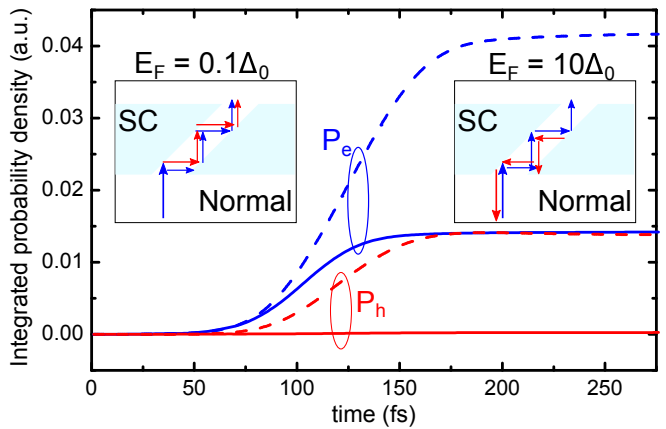}}
\caption{(Color online) Electron (blue) and hole (red) integrated probability densities as a function of time in the normal region beyond a $W = 300$ \AA\,, $L = 300$ \AA\, channel in the superconducting region, assuming Fermi energies $E_F = 10 \Delta_0$ (solid) and $E_F = 0.1 \Delta_0$ (dashed). Arrows in the insets illustrate the expected electron and hole trajectories undergoing reflections by the normal-SC interfaces.} \label{fig:DeviceTrajectories}
\end{figure}

The time evolution of $P_e$ (blue) and $P_h$ (red), integrated in the region after the SC ($[y_1, y_2] = [L/2,\infty]$ and $[x_1, x_2] = [-\infty,\infty]$, see Fig. \ref{fig:Sketch}(b)) is shown in Fig. \ref{fig:DeviceTrajectories}, assuming $E_F = 10 \Delta_0$ (solid) and $E_F = 0.1 \Delta_0$ (dashed). In general, all $P_e$ and $P_h$ values are small, due to the fact that most of the incoming electron wave front reaches the SC region aside of the channel entrance, and just a small fraction of it is actually capable of entering the channel region. The probability of finding an electron after the SC region is always non-zero, and it is higher for $E_F = 0.1 \Delta_0$. However, the probability for holes to cross the channel is non-zero only for $E_F = 0.1 \Delta_0$. The trajectories of electrons (blue) and holes (red) illustrated in the insets help to understand this feature. As the electron is horizontally (vertically) reflected by the first (second) normal-SC interface in the channel, the resulting holes propagate in a direction that depends on $E_F$. For $E_F > \Delta_0$, the retro-reflected holes created in each normal-SC reflection propagate backwards along the same trajectory of the ongoing electron, thus, no hole is able to cross the channel. Conversely, for $E_F < \Delta_0$, specular-reflected holes arisen in each normal-SC reflection propagate along with the electron across the channel and eventually make their way through it, thus yielding non-zero hole probability beyond the channel.

\begin{figure}[!t]
\centerline{\includegraphics[width = 0.9\linewidth]{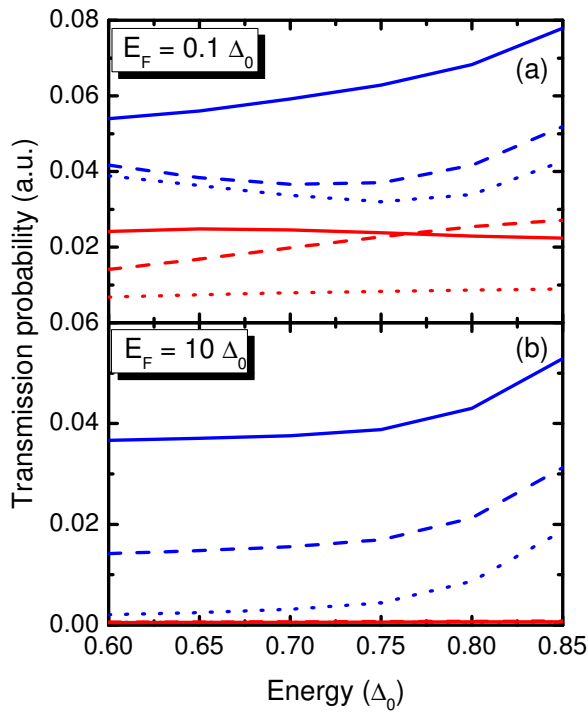}}
\caption{(Color online) Electron (blue) and hole (red) integrated transmission probability as a function of the energy of the incoming wave front, assuming a channel in the superconducting region with length $L = 300$ \AA\,, assuming widths $W = 200$ \AA\, (dotted), 300 \AA\, (dashed), and 400 \AA\, (solid). Fermi energies are (a) $E_F = 0.1 \Delta_0$ and (b) $E_F = 10 \Delta_0$.} \label{fig:DeviceResultsW}
\end{figure}

\begin{figure}[!t]
\centerline{\includegraphics[width = 0.9\linewidth]{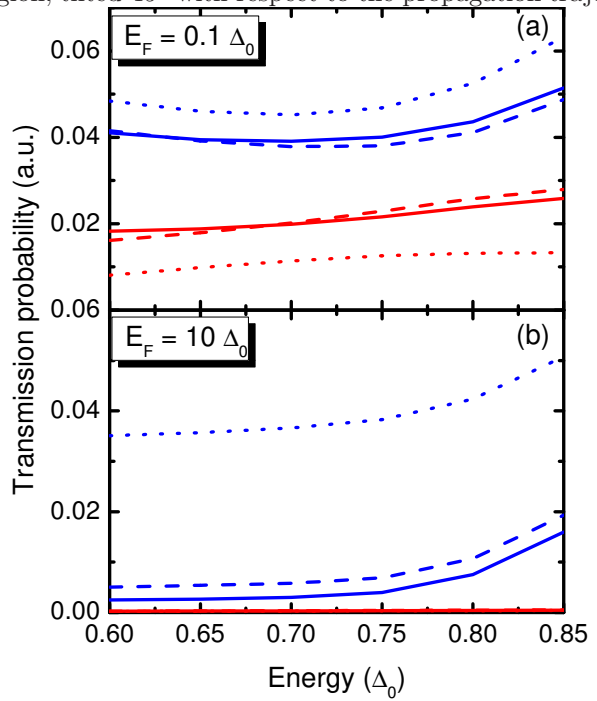}}
\caption{(Color online) Electron (blue) and hole (red) integrated transmission probability as a function of the energy of the incoming wave front, assuming a channel in the superconducting region with width $W = 300$ \AA\,, for lengths $L = 200$ \AA\, (dotted), 350 \AA\, (dashed), and 400 \AA\, (solid). Fermi energies are (a) $E_F = 0.1 \Delta_0$ and (b) $E_F = 10 \Delta_0$.} \label{fig:DeviceResultsL}
\end{figure}

The dependence of the electron and hole transmission probabilities on the width $W$ and length $L$ of the channel is shown in Figs. \ref{fig:DeviceResultsW} and \ref{fig:DeviceResultsL}, respectivelly. For $E_F = 0.1 \Delta_0$ and a fixed length $L = 300$ \AA\,, results in Fig. \ref{fig:DeviceResultsW}(a) show that increasing the channel width $W$ from 200 \AA\, to 400 \AA\, improves the hole transmission probability for wave-packet energies lower than $\approx 0.775 \Delta_0$. For higher energies, hole transmission probability for $W = 300$ \AA\, is just slightly lower thant that for $W = 400$ \AA\,. Nevertheless, a significant hole transmission probability is observed only for $E_F = 0.1 \Delta_0$. For $E_F = 10 \Delta_0$, Fig. \ref{fig:DeviceResultsW}(b) show an electron transmission probability that monotonically increase with the wave-packet energy, whereas hole probabilities are always vanishingly small. Qualitatively, this result persists for the whole energy range considered here, namely from $\varepsilon$ = 0.6 $\Delta_0$ to $\varepsilon$ = 0.85 $\Delta_0$. Similar conclusions are also drawn from the results in Fig. \ref{fig:DeviceResultsL}, where increasing the channel length $L$ is demonstrated to yield equivalent results as decreasing the width $W$.

The 45$^o$ value was chosen for the angle of the tilted channel only for convenience, in order to facilitate the visualization of the results. One can easily verify that the same qualitative results would be observed for any angle. In fact, even a straight vertical channel shows a non-zero transmission probability for holes in the $E_F < \Delta_0$ case. However, this effect is much weaker for a vertical channel, since electron-hole conversion requires the wave function to bounce back and forth between the normal-SC interfaces in the channel, which is optimized as the angle between the channel and the vertical axis increase.      

\section{Conclusions}

In summary, we have proposed a general numerical technique to investigate electron scattering and electron-hole conversion at normal-SC interfaces with arbitrary shapes and profiles. The method, based on real time wave-packet propagation through a system described by a Bogoliubov-de Gennes model, is easily adapted for Hamiltonians representing different materials, and allows one to observe electron and hole trajectories in a pedagogical and convenient way. As a sample case, we apply the method to revisit the problem of Andreev reflection in a normal-SC interface in monolayer graphene, where the transition from retro-reflection to specular reflection is observed just by tracking electron and hole trajectories as the Fermi level of the system is tuned. 

As an example of an arbitrary profile of the SC region, we consider the case of an electron wave front propagating through a normal channel within the superconducting region, tilted 45$^o$ with respect to the propagation trajectory of the incoming electron. The system is demonstrated to work as an electronic wave guide for any value of Fermi level $E_F$. However, the channel guides holes along with the electrons only for $E_F < \Delta_0$, whereas the retro-reflected holes in the $E_F > \Delta_0$ case propagate backwards and leave the channel via its entrance. This effect is enhanced as either the channel length or width are increased.

Exciting future prospects for this method are to investigate Andreev reflection in e.g. multi-layer graphene, monolayer transition metal dichalcogenides, and phosphorene, even under external applied electric and/or magnetic fields. Required modifications are straightforward, and therefore expected in imminent following studies.     

\acknowledgements This work was supported by the Brazilian Council for Research (CNPq), through the PRONEX/FUNCAP and PQ programs, and by the Flemish Research Foundation (FWO).

\end{document}